\documentclass[letterpaper, 10 pt, conference]{ieeeconf}      
\usepackage{booktabs}
\usepackage{graphicx}
\usepackage{cite}
\usepackage{url}
\usepackage{algorithm}
\usepackage{algorithmic}
\usepackage{amsmath}
\usepackage{caption}
\usepackage{subcaption}
\usepackage{amssymb}
\usepackage{booktabs}
\usepackage{graphicx}
\usepackage{cite}
\usepackage{url}
\usepackage{algorithm}
\usepackage{algorithmic}
\usepackage{amsmath}
\usepackage{caption}
\usepackage{subcaption}
\usepackage{bm}
\usepackage{bbm}
\usepackage{amsfonts}
\usepackage{float}
\usepackage{multirow}
\usepackage{threeparttable}
\usepackage{dcolumn}
\usepackage{multicol}
\usepackage{blindtext}
\usepackage{subcaption}
\usepackage{color,soul}
\usepackage{units,color}
\usepackage{pbox}
\usepackage{epstopdf}

\newtheorem{problem}{Problem}

\graphicspath{ {figs/}}

\IEEEoverridecommandlockouts                              

\title{\LARGE \bf
	Ocean Plume Tracking with Unmanned Surface Vessels: Algorithms and Experiments}
\author{Muhammad~Fahad$^{1}$,
		Yi~Guo$^{1}$,
		Brian~Bingham$^{2}$,
		Kristopher~Krasnosky$^{3}$,
		Laura~Fitzpatrick$^{3}$,\\
		and~Fernando~A.~Sanabria$^{3}$
\thanks{The work was partially supported by the National Science Foundation under Grants IIS-1218155 for Stevens and IIS-1217659 for UH.}%
\thanks{$^{1}$Muhammad Fahad and Yi Guo are with Department of Electrical \& Computer Engineering, Stevens Institute of Technology, Hoboken, NJ 07030, USA.
	{\tt\small mfahad, yguo1@stevens.edu}}%
\thanks{$^{2}$ Brian Bingham is with the Department of Mechanical and Aerospace Engineering, Naval Postgraduate School,
	Monterey, CA 93950, USA.
	{\tt\small bbingham@nps.edu}}%
\thanks{$^{3}$ Kristopher~Krasnosky, Laura~Fitzpatrick, and Fernando~A.~Sanabria are with the Department of Mechanical Engineering, University of Hawaii at Manoa,
	Honolulu, HI 96822, USA
	{\tt\small kekrasno, laura24, faragon@hawaii.edu}}%
}

\begin{document}

	\maketitle
	\thispagestyle{empty}
	\pagestyle{empty}
\begin{abstract}

Pollution plume monitoring using autonomous vehicles is important due to the adverse effect of pollution plumes on the environment and associated monetary losses. Using the advection-diffusion plume dispersion model, we present a control law design to track dynamic concentration level curves. We also present a gradient and divergence estimation method to enable this control law from concentration measurement only. We then present the field testing results of the control law to track concentration level curves in a plume generated using Rhodamine dye as a pollution surrogate in a near-shore marine environment. These plumes are then autonomously tracked using an unmanned surface vessel equipped with fluorometer sensors. Field experimental results are shown to evaluate the performance of the controller, and complexities of field experiments in real-world marine environments are discussed in the paper.

\end{abstract}

\section{Introduction}\label{sec:intro}

Pollution plumes not only cause short-term and long-term damage to the environment but also have adverse societal impacts that manifest as economic loss and health impacts for people living in the affected area. These effects are most clearly observable for oil plumes after oil spills in marine environments \cite{chang2014consequences}. Tracking pollution plumes has historically relied on plume trajectory models and remote sensing platforms. Both these methods are limited in effectiveness \cite{klemas2010tracking}. Modeling techniques require very accurate environmental data and leak source parameters. On the other hand, remote sensing solutions are expensive and limited by weather conditions, and can only track surface plumes. With the development of long endurance autonomous unmanned vessels (AUV), using these to track the spatial extent of pollution is an appealing solution that resolves some of these problems.

Using robots for environmental monitoring has previously been studied mainly for source seeking, where robots are employed for locating the source of a chemical plume in both aerial and marine domains. A thorough review of existing work in these domains was provided by Dunbabin and Marques in \cite{dunbabin2012robots}, which classifies existing techniques into gradient based, biologically inspired, and stochastic search methods. Another review by Ishida et. al.\cite{6247459} summarizes the use of robotic platforms for monitoring chemicals introduced in a fluid medium. Field experiment results for chemical source seeking have been presented in \cite{farrell2005chemical,1283418}, and plume mapping experiments have been presented in \cite{lilienthal2004building,ferri2011mapping}.

Level curve tracking is related to research directions of gradient climbing\cite{ogren2004cooperative,5398831}, estimating environmental boundaries \cite{bertozzi2005determining,hsieh2005experimental}, perimeter patrolling \cite{clark2007mobile,shuailipaper}, and sample coverage of a large area \cite{smith2011persistent,ferri2011mapping}. Zhang and Leonard in \cite{5398831} presented a method to track static level curves using a gradient based approach, where gradient information was used to minimize the square error between the robot location and the location of the level curve. Hseih et. al. in \cite{hsieh2005experimental} developed a boundary tracking algorithm that uses simple circular motion as a building block for a composite path that eventually travels the entire boundary of the region. Clark and Fierro in \cite{clark2007mobile} presented a biological behavior mimicking system to patrol both static and dynamic environmental perimeters. Persistent ocean monitoring experiments were presented in \cite{smith2011persistent}, where underwater gliders swept to perform sample coverage of an area using lawn mower paths to study ocean phenomenon of the occurrence and life cycle of harmful algal blooms. However, most algorithm categories consider {\em static} level curves that are temporally non-evolving with a few exceptions \cite{clark2007mobile}. 

While previous studies demonstrate either sources seeking strategies or plume mapping by exhaustive search methods, our approach addresses the problem of tracking dynamic level curves. Level curve tracking has recently received increased research attention. In this work, we study control based methods for {\em dynamic} level curve tracking based on an advection-diffusion pollution dispersion model. We also address the problem of concentration gradient and divergence estimation from point measurements. We present an estimation method of these quantities using available point concentration measurements, and then test the performance of the tracking control algorithms in real field experiments using an unmanned surface vessel (USV) and Rhodamine dye plumes.

\section{The Model and Problem Statement}\label{sec:algorithm}

In this section we first present the kinematic model of the USV used in this work. We then provide the details of the pollutant plume propagation model and followed by a formal description of the control objective.

\subsection{Vessel Model}

The kinematic model of the USV used in these experiments is described by
\begin{eqnarray}\label{eqn3}
\dot{x}& =& \nu_r\cos{\theta_r} \nonumber,
\\
\dot{y}&=& \nu_r\sin{\theta_r},
\\
\dot{\theta}_r&=&\omega_r \nonumber,
\end{eqnarray}
where $\bm{\mathrm{x}}_r=[x, y]^T$ represents the USV location, $\theta_r$ is the USV orientation and $\nu_r, \omega_r$ are the translational and rotational velocities, respectively. We define a new variable $z$ as $z=[z_1,z_2]^T=[x+l_0\cos{\phi},y+l_0\sin{\phi}]^T$ where $l_0$ is a positive constant. Then (\ref{eqn3}) can be re-written as a single integrator model
\begin{equation}\label{eqn4}
\dot{z}=u_r,
\end{equation}
where the new control input $u_r=[u_1, u_2]^T$ is defined as
\begin{equation}\label{eqn5}
u_r= \left[
  \begin{array}{cc}
    \cos{\theta_r} & -l_0\sin{\theta_r} \\
    \sin{\theta_r} & l_0\cos{\theta_r}
  \end{array}
 \right] \cdot\left[
        \begin{array}{cc}
        \nu_r\\
        \omega_r
         \end{array}
 \right] \stackrel{\triangle}{=} C\cdot \left[
        \begin{array}{cc}
        \nu_r\\
        \omega_r
         \end{array}
 \right].
\end{equation}
Note that $C$ is invertible, and its inverse is given by D=$\left[
  \begin{array}{cc}
    \cos{\theta_r} & \sin{\theta_r} \\
    \frac{-\sin{\theta_r}}{l_0} & \frac{-\cos{\theta_r}}{l_0}
  \end{array}
 \right]$.
In the USV control design, we first design the control input $u_r$ for the integrator model in (\ref{eqn5}), and then the original vehicle input  $\nu_r, \omega_r$ can be obtained by the inverse operation of (\ref{eqn5}).

\subsection{Plume Propagation Model}\label{sec:plume_propogation_model}

The spatiotemporal growth of the plume can be modeled by two mechanisms, advection and diffusion. Advection is the spread of the plume due to flow of the fluid, and diffusion is its spread due to its motion from  higher concentration to lower concentration. The partial differential equation describing this spread in two dimensions can be written as

\begin{equation}\label{eqn1}
\begin{split}
\frac{\partial c(\bm{\mathrm{x}},t)}{\partial t}+\bm{\mathrm{v}}^T(\bm{\mathrm{x}},t)\nabla c(\bm{\mathrm{x}},t) = k \nabla^2 c(\bm{\mathrm{x}},t),
\end{split}
\end{equation}
where $c(\bm{\mathrm{x}},t)$ is the concentration at spatial location $\bm{\mathrm{x}}$, $\bm{\mathrm{v}}(\bm{\mathrm{x}},t)$ is the fluid flow field vector, $\nabla c(\bm{\mathrm{x}},t)=\frac { \partial c(\bm{\mathrm{x}},t)}{\partial \bm{\mathrm{x}}}$ is the spatial gradient of $c(\bm{\mathrm{x}},t)$, $\nabla^2 c(\bm{\mathrm{x}},t)=\frac { \partial^2 c(\bm{\mathrm{x}},t)}{\partial x^2}+\frac { \partial^2 c(\bm{\mathrm{x}},t)}{\partial y^2}$ is the divergence of $c(\bm{\mathrm{x}},t)$ in two dimensional space  $\bm{\mathrm{x}}=[x,y]^T$, $k$ is the turbulent diffusion coefficient. The concentration field in (\ref{eqn1}) is a time varying quantity. 

A set of spatial locations, that have the same concentration value c($\bm{\mathrm{x}},t$) = c$_0$ (here c$_0>0$ is the monitored concentration) is defined as a level curve in this concentration field. This set of spatial locations can be formally defined as the concentration level set: \{$\bm{\mathrm{x}}\in\mathbb{R}^2$,c($\bm{\mathrm{x}},t$) = c$_0$\}. 
\subsection{Problem Statement}\label{sec:plume_monitoring}
To monitor the pollutant plume described in (\ref{eqn1}) using a USV, we assume that the USV has on-board sensors including: 1) localization sensors to obtain its position  $\bm{\mathrm{x}}_r$ and heading $\theta_r$, 2) concentration measurement sensors to measure the plume concentration $c_r$ at the robot position $\bm{\mathrm{x}}_r$, and 3) acoustic Doppler current profilers to measure the flow velocity $\bm{\mathrm{v}}_r$ at the robot position $\bm{\mathrm{x}}_r$. The control objective is to drive the USV along the concentration level curve. In addition, we add a second control objective to patrol along the level curve with a desired speed v$_d$. The control objectives are formally stated as follows.
\begin{problem}\label{problem:statement}
For the USV modeled by (\ref{eqn3}), design a control law to drive the USV to track the concentration level curve \{$\bm{\mathrm{x}}\in\mathbb{R}^2$,c($\bm{\mathrm{x}},t$) = c$_0$\}, and patrol along it with a desired speed $\mathrm{v}_d$.
\end{problem}

\section{The Control Algorithm and Gradient Estimation}\label{sec:control_algorithm}
In this section we present the design of the control law to solve the plume monitoring problem stated in Problem~\ref{problem:statement}. This control law is inspired by the plume front monitoring controller described in our previous work \cite{shuailipaper}, but we remove the assumption that the gradient information is available for controller use. 
\subsection{Control Algorithm}
The level curve tracking task can be divided into two parts. First the estimation part estimates the location of the level curve. The second part is the tracking control part, which drives the USV to the estimated level curve. Let $\bm{\mathrm{x}}$ represent the location of the level concentration curve with concentration $c_0$ at time $t$. The estimator estimates $\hat{\bm{\mathrm{x}}}$, the location of this level concentration curve. The estimator is designed to enable the convergence of $\hat{\bm{\mathrm{x}}}$ to $\bm{\mathrm{x}}$, i.e., the estimation error $\bm{\mathrm{e}}$~=~$\bm{\mathrm{x}}$-$\hat{\bm{\mathrm{x}}}$ converges to zero.

The controller design follows the method proposed in \cite{shuailipaper}. Considering the level concentration curve, $c(\bm{\mathrm{x}},t)=c_0$, and using the derivative of the concentration curve and the model presented in (\ref{eqn1}), the plume front dynamics can be written as,
\begin{equation}\label{eqn:level_cur}
\dot{\bm{\mathrm{x}}}^T \nabla c = -\bm{\mathrm{v_x}}^T \nabla c - k\nabla^2c
\end{equation}
where $\bm{\mathrm{v_x}}$=$\bm{\mathrm{v}}(\bm{\mathrm{x}},t)$, $\nabla c =\nabla c(\bm{\mathrm{x}},t)$, and $\nabla^2 c =\nabla^2 c(\bm{\mathrm{x}},t)$. The time $t$ has been dropped in (\ref{eqn:level_cur}) and in the subsequent discussion without causing confusion. Since the USV is required to patrol along the concentration curve with velocity v$_d$, additional behavior constraint can be described as
\begin{equation}\label{eqn:patrolling}
\frac{(A\nabla c)^T}{\|\nabla c\|}\dot{\bm{\mathrm{x}}}=\mathrm{v}_d,
\end{equation}
where A $=\left[
\begin{array}{cc}
0 & -1 \\
1 & 0
\end{array}
\right]$ is an orthogonal rotation matrix.
The estimated level curve location can now be calculated as
\begin{equation}\label{eqn:level_curve_estimation}
\dot{\bm{\mathrm{x}}}=-\frac{(\bm{\mathrm{v_x}}^T \nabla c - k \nabla^2c)\nabla c}{\|\nabla c\|^2}+\frac{(\mathrm{v}_d A\nabla c)}{\|\nabla c\|}.
\end{equation}
The measurement at the observed level curve location $\hat{\bm{\mathrm{x}}}$, using the USV local measurement and first order Taylor series expansion, can be given by
\begin{equation}\label{eqn:measurement}
c_{\hat{\bm{\mathrm{x}}}}=\nabla^T c_r(\hat{\bm{\mathrm{x}}}-\bm{\mathrm{x}}_r)+c_r,
\end{equation}
where $\bm{\mathrm{x}}_r$ is the USV location, $c_r$ = $c(\bm{\mathrm{x}}_r,t)$, $c_{\hat{\bm{\mathrm{x}}}}$ = $c(\hat{\bm{\mathrm{x}}},t)$ and $\nabla c_r = \nabla c(\bm{\mathrm{x}}_r,t)$.

Using (\ref{eqn:level_curve_estimation}) and (\ref{eqn:measurement}), the following Luenberger state observer \cite{zeitz1987extended} can be used:
\begin{eqnarray}\label{eqn:oberserver_not_robot}
\dot{\hat{\bm{\mathrm{x}}}}&=&-\frac{({\bm{\mathrm{v}}_{\hat{\bm{\mathrm{x}}}}}^T \nabla c_{\hat{\bm{\mathrm{x}}}}+k \nabla^2 c_{\hat{\bm{\mathrm{x}}}})\nabla c_{\hat{\bm{\mathrm{x}}}}}{\|\nabla c_{\hat{\bm{\mathrm{x}}}}\|^2}+ \frac{ \mathrm{v}_d A\nabla c_{\hat{\bm{\mathrm{x}}}} }{\|A\nabla c_{\hat{\bm{\mathrm{x}}}}\|} \nonumber\\
&&-k_1 \nabla c_{\hat{\bm{\mathrm{x}}}} \big(\nabla^T c_{\hat{\bm{\mathrm{x}}}}(\hat{\bm{\mathrm{x}}}-{\bm{\mathrm{x}}}_r)+c_r-c_0\big),
\end{eqnarray}
where $k_1>0$ is a coefficient, ${\bm{\mathrm{v}}_{\hat{\bm{\mathrm{x}}}}}$ = $\bm{\mathrm{v}}(\hat{\bm{\mathrm{x}}},t)$, $\nabla c_{\hat{\bm{\mathrm{x}}}}$ = $\nabla c(\hat{\bm{\mathrm{x}}},t)$ and $\nabla^2 c_{\hat{\bm{\mathrm{x}}}}$ = $\nabla^2 c(\hat{\bm{\mathrm{x}}},t)$.
Since the quantities are immeasurable at $\hat{\bm{\mathrm{x}}}$, we simply replace these with the corresponding quantities at the USV location $\bm{\mathrm{x}}_r$ and rewrite the equation as
\begin{eqnarray}\label{eqn6}
\dot{\hat{\bm{\mathrm{x}}}}&=&-\frac{({\bm{\mathrm{v}}_r}^T \nabla c_r+k \nabla^2 c_r)\nabla c_r}{\|\nabla c_r\|^2}+ \frac{ \mathrm{v}_d A\nabla c_r }{\|A\nabla c_r\|} \nonumber\\
&&-k_1 \nabla c_r \big(\nabla^T c_r(\hat{\bm{\mathrm{x}}}-\bm{\mathrm{x}}_r)+c_r-c_0\big),
\end{eqnarray}
where  $\bm{\mathrm{v}}_r$ = $\bm{\mathrm{v}}(\bm{\mathrm{x}}_r,t)$, $\nabla c_r$ = $\nabla c(\bm{\mathrm{x}}_r,t)$ and $\nabla^2 c_r$ = $\nabla^2 c(\bm{\mathrm{x}}_r,t)$.

The control input $u_r$ for the USV to track the estimated plume front $\hat{\bm{\mathrm{x}}}$ is designed as follows:
\begin{eqnarray}\label{eqn7}
u_r&=&-\frac{({\bm{\mathrm{v}}_r}^T \nabla c_r+k \nabla^2 c_r)\nabla c_r}{\|\nabla c_r\|^2}+ \frac{ \mathrm{v}_d  A\nabla c_r }{\|A\nabla c_r\|}-k_1 \nabla c_r \nonumber\\
&&\cdot\big(\nabla^T c_r(\hat{\bm{\mathrm{x}}}-\bm{\mathrm{x}}_r)+c_r-c_0\big)-k_2(\bm{\mathrm{x}}_r- \hat{\bm{\mathrm{x}}}),
\end{eqnarray}
where $k$, $k_1$, $k_2$ are positive constants representing the diffusion coefficient, gradient gain, and tracking gain, respectively. Following the same convergence analysis as shown in \cite{shuailipaper}, it can be proved that this controller makes the estimation error $\bm{\mathrm{e}}$~=~$\bm{\mathrm{x}}$-$\hat{\bm{\mathrm{x}}}$ go to zero as time elapses. Note that this controller makes the USV reach the concentration level curve and patrol along the level curve in a counter-clock direction (due to the second term in (\ref{eqn6})), thus solve Problem 1 defined in Section~\ref{sec:algorithm}.

Since the sensors installed on the USV are point sensors and can only measure the concentration value at one point, the concentration gradient $\nabla c_r$ and divergence $\nabla^2 c_r$ are not directly available. 
\subsection{Gradient Estimation}\label{sec:grad_est}

\begin{figure}[h]
	\centering
	\includegraphics[width=0.15\textwidth,angle =0]{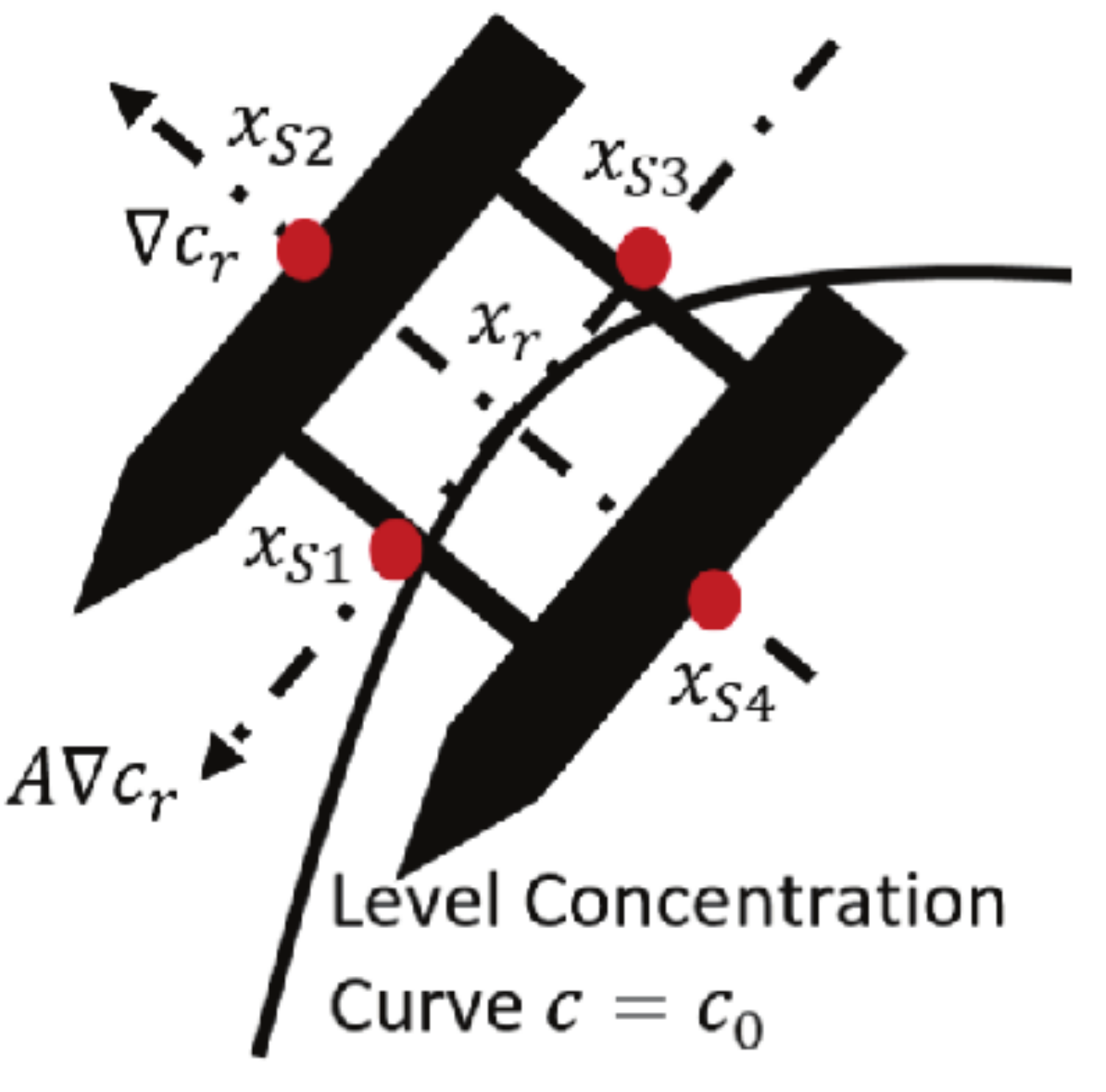}
	\caption{Sensor configuration for the USV. $\bm{\mathrm{x}}_{S1}$, $\bm{\mathrm{x}}_{S2}$, $\bm{\mathrm{x}}_{S3}$ and $\bm{\mathrm{x}}_{S4}$ denote the location of the four sensors on the USV, marked by the red circles. The USV center is denoted by $\bm{\mathrm{x}}_r$. }
	\label{sensor_location}
\end{figure}
The gradient $\nabla c_r$ and divergence $\nabla^2 c_r$ at the USV location $\bm{\mathrm{x}}_r$ is needed to generate the control input in (\ref{eqn7}). The sensors installed on the USV are point sensors, and can only measure concentration levels at one spatial location. Therefore $\nabla c_r$, and $\nabla^2 c_r$ must be estimated from these point measurements. The estimation method for $\nabla c_r$ and $\nabla^2 c_r$ is discussed in this section. The sensor locations configured on the USV are shown in Fig.~\ref{sensor_location}. Let $\bm{\mathrm{x}}_{Si}$ represent the location of the $i$-th sensor, $i$=1,2,3,4, on the USV. Given that the sensors are sufficiently close to the USV center $\bm{\mathrm{x}}_r$, the concentration $c(\bm{\mathrm{x}}_{Si})$ at the sensor location can be approximated by a second-order Taylor approximation, i.e.,
\begin{eqnarray}\label{2015-04-20-1}
c(\bm{\mathrm{x}}_{Si})&\approx& c_r+(\bm{\mathrm{x}}_{Si}-\bm{\mathrm{x}}_{r})^T\nabla c_r\nonumber\\
&+&0.5(\bm{\mathrm{x}}_{Si}-\bm{\mathrm{x}}_{r})^T H (\bm{\mathrm{x}}_{Si}-\bm{\mathrm{x}}_{r}),
\end{eqnarray}
for $i\in\{1,2,3,4\}$, where $H$ is the Hessian matrix. For the matrix $H$, the notation $\textit{\textbf{h}}$ is used to represent a column vector defined by rearranging the elements of $H$ as $\textit{\textbf{h}}$$=[H_{11},H_{12},H_{21},H_{22}]^T$. The gradient and divergence estimation using these equations are given by \cite{briggs2011calculus} as $\boldsymbol{\gamma}=\textit{\textbf{B}}^+\textit{\textbf{y}},$ where $\textit{\textbf{B}}^+$ is the Moore-Penrose pseudoinverse of $\textit{\textbf{B}}$ defined as $\textit{\textbf{B}}^+=\textit{\textbf{B}}^T(\textit{\textbf{B}}\textit{\textbf{B}}^T)^{-1}$, and
\begin{eqnarray}
\nonumber
\boldsymbol{\gamma}&=&[\nabla^T c(\bm{\mathrm{x}}_{r}),\textit{\textbf{h}}^T]^T,\\
\textit{\textbf{y}}&=&\left[
                                             \begin{array}{c}
                                               c(\bm{\mathrm{x}}_{S1})-\widehat{c_r}\\
                                               c(\bm{\mathrm{x}}_{S2})-\widehat{c_r} \\
                                               c(\bm{\mathrm{x}}_{S3})-\widehat{c_r}\\
                                               c(\bm{\mathrm{x}}_{S4})-\widehat{c_r}\\
                                             \end{array}
                                           \right],\\
\textit{\textbf{B}}&=&\left[\begin{array}{cc}
              (\bm{\mathrm{x}}_{S1}-\bm{\mathrm{x}}_{r})^T & \ \\
              (\bm{\mathrm{x}}_{S2}-\bm{\mathrm{x}}_{r})^T & \textit{\textbf{E}} \\
              (\bm{\mathrm{x}}_{S3}-\bm{\mathrm{x}}_{r})^T &\ \\
              (\bm{\mathrm{x}}_{S4}-\bm{\mathrm{x}}_{r})^T &\ \\
            \end{array}
          \right]
\end{eqnarray}
with
$\bm{\mathrm{x}}_{r}=0.25\sum_{i=1}^{4}\bm{\mathrm{x}}_{Si},\
\widehat{c_r}=0.25\sum_{i=1}^{4}c(\bm{\mathrm{x}}_{Si}),$ and
$$\textit{\textbf{E}}=0.5\left[
                        \begin{array}{c}
                          \overrightarrow{(\bm{\mathrm{x}}_{S1}-\bm{\mathrm{x}}_{r})(\bm{\mathrm{x}}_{S1}-\bm{\mathrm{x}}_{r})^T} \\
                          \overrightarrow{(\bm{\mathrm{x}}_{S2}-\bm{\mathrm{x}}_{r})(\bm{\mathrm{x}}_{S2}-\bm{\mathrm{x}}_{r})^T} \\
                          \overrightarrow{(\bm{\mathrm{x}}_{S3}-\bm{\mathrm{x}}_{r})(\bm{\mathrm{x}}_{S3}-\bm{\mathrm{x}}_{r})^T} \\
                          \overrightarrow{(\bm{\mathrm{x}}_{S4}-\bm{\mathrm{x}}_{r})(\bm{\mathrm{x}}_{S4}-\bm{\mathrm{x}}_{r})^T} \\
                        \end{array}
                      \right]
.$$

Hence, the estimates $\widehat{c_r}$, $\widehat{\nabla c_r}$ and $\widehat{\nabla^2 c_r}$ can be obtained as

$$\widehat{c_r}=0.25\sum_{i=1}^{4}c(\bm{\mathrm{x}}_{Si}),$$

$$\widehat{\nabla c_r}=\left[
                                                     \begin{array}{cccccc}
                                                       1 & 0 & 0 & 0 & 0 & 0 \\
                                                       0 & 1 & 0 & 0 & 0 & 0 \\
                                                     \end{array}
                                                   \right]\boldsymbol{\gamma},$$

$$\widehat{\nabla^2 c_r}=\left[
                                                     \begin{array}{cccccc}
                                                       0 & 0 & 1 & 0 & 0 & 1 \\
                                                                                                           \end{array}
                                                   \right]\boldsymbol{\gamma}.$$

\begin{figure}[h]
	\centering
	\includegraphics[width=0.25\textwidth,angle =0]{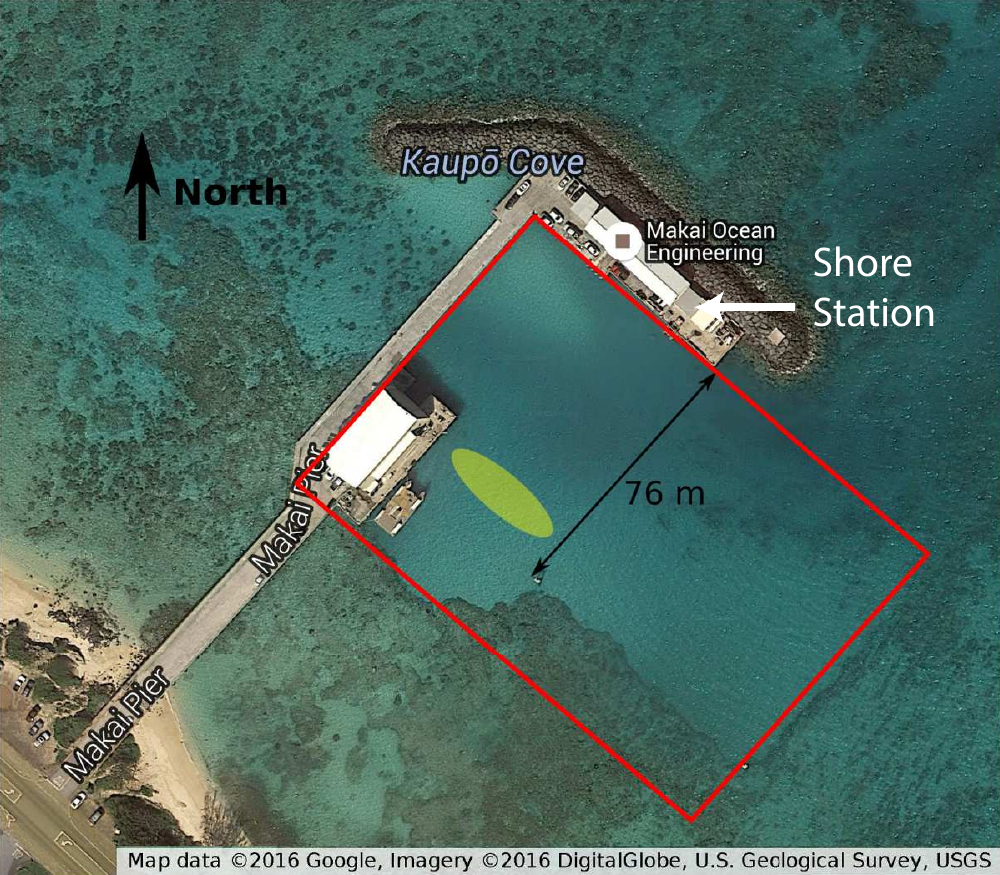}
	\caption{Field Experiment Site at Makai Research Pier Oahu, Hawaii. The ellipse marks the approximate source location during different experiments and red rectangle marks the testing arena.}
	\label{makai_field_final}
\end{figure}

\section{Field Experiments}\label{sec:experiment_results}
In this section, we present the setup used to perform the experiments and the results of these experiments followed by a discussion.
\subsection{Experiment Setup}
The experiments were carried out at the Makai Research Pier in August 2015. The testing site was chosen due to its relatively calmer ocean conditions. The testing site is sheltered by a pier and breakwater on two sides, the shore on one side and the fourth side is open to the sea. The detailed testing site geography is shown in Fig.~\ref{makai_field_final}. These conditions provided relatively calmer near shore ocean conditions for testing. The ocean floor at the site is relatively flat and sandy. The average depth of the sea floor here is approximately 4~m \cite{test_site}. The experiments were conducted in an arena approximately marked by the red boundary in Fig.~\ref{makai_field_final}. The approximate location of the plume source is also marked by an ellipse in this figure. An inflatable raft with plume generating equipment was used as the plume source. The raft had a constant flow rate pump onboard that dosed Rhodamine dye into the ocean to generate the plume. Rhodamine dye is a harmless substance, disperses over time and was used as a surrogate to generate plumes during these experiments.
\begin{figure}[tbph]
	\centering
	\includegraphics[width=0.25\textwidth,angle =0]{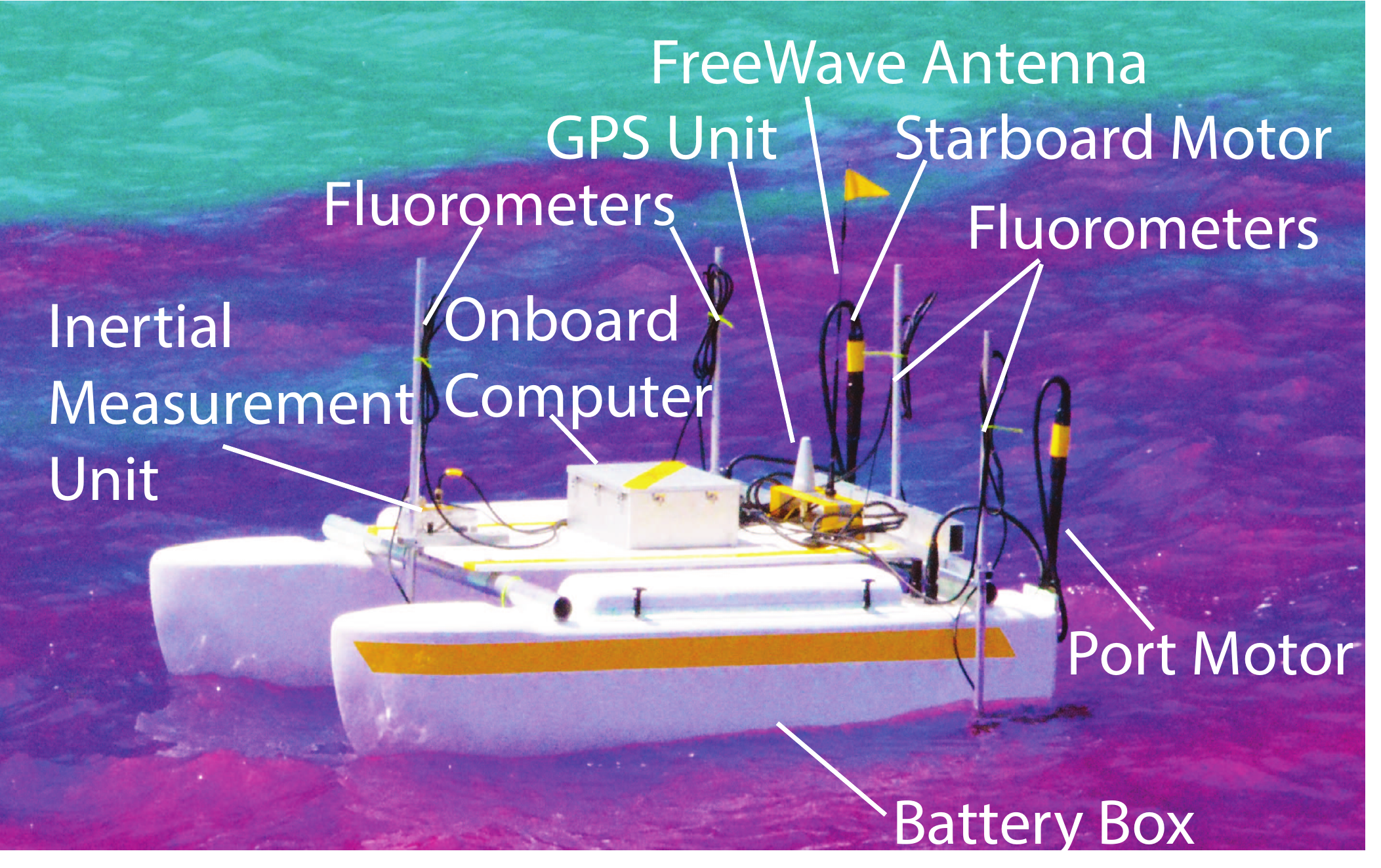}
	\caption{Unmanned surface vessel (USV) developed by Field Robotics Lab at University of Hawaii. The pink color seen around the USV is Rhodamine dye.}
	\label{UPSV}
\end{figure}
The USV platform used in this work was originally developed at the University of Hawaii for port and harbor security applications \cite{bingham11development}. The original vessel was modified to improve the design and to include additional sensors required for the current experiments. This is a twin-hull catamaran type vessels, equipped with dual starboard and port actuation. The USV used in the experiments illustrated in Fig.~\ref{UPSV}, is 2~m long, 1.5~m wide, weighs approximately 150~lbs and can carry a payload of 180~lbs. It is capable of operating at 1~m/s nominal speed, 2~m/s maximum speed, and can operate at the nominal speed for 4~hrs on a single charge. 

The fluorometer sensors were installed on the USV to measure the concentration of the dye in the ocean. Cyclops~7 fluorometers by Turner Design were used for this purpose. The USV was designed to support up to a maximum of four sensors. The minimum detection threshold of this sensor is 0.01~ppb and its dynamic range is 0-10,000~ppb~\cite{man:turnderdesign}. To ensure accuracy of the measured concentration data from the sensors, the fluorometers used in the experiments require precise calibration in a laboratory setting before they can be used in experiments. The calibration was performed by mixing known Rhodamine dye concentration with known quantity of sea water, creating a solution whose concentration was known. The concentration of the resultant mixture was measured with the fluorometers and calibration records were prepared. 

Each experiment was thus started with establishing the expected direction of the plume growth. This was done by releasing a small amount of dye in the ocean and observing its advection direction. Once this was established, the plume source was placed approximately in the oval area, highlighted in Fig.~\ref{makai_field_final}, to ensure maximum plume growth in the designated area, without interaction with the pier. The tests were conducted using the USV configured to carry four fluorometer sensors in the sensor configuration detailed in Fig.~\ref{sensor_location}. 

\subsection{Experiment Results}
Two different level concentration curves were tracked during these tests, first case to track 50~ppb and second case to track 40~ppb. The results for these different tracking concentrations are detailed next.

\subsubsection{Case 1}$\textit{Tracking level concentration curve at 50~ppb}$
The first test was conducted with the tracked concentration $c_0$ set to 50~ppb. This was done to test the performance of the controller to track a concentration value in the denser section of the plume. The constants $k$, $k_1$ and $k_2$ in (\ref{eqn7}) were set to 1.2, 5 and 11 respectively. The desired patrolling speed v$_d$ was set to 1.5~m/s. The test was conducted for 60~secs. The USV stayed within the denser section of the plume, generating a near circular trajectory. 

\begin{figure*}[h]
	\centering
	\captionsetup{justification=centering}
	\begin{subfigure}[t]{0.25\textwidth}
		\centering
		\includegraphics[width=0.8\textwidth, height=0.8\textwidth, angle =0]{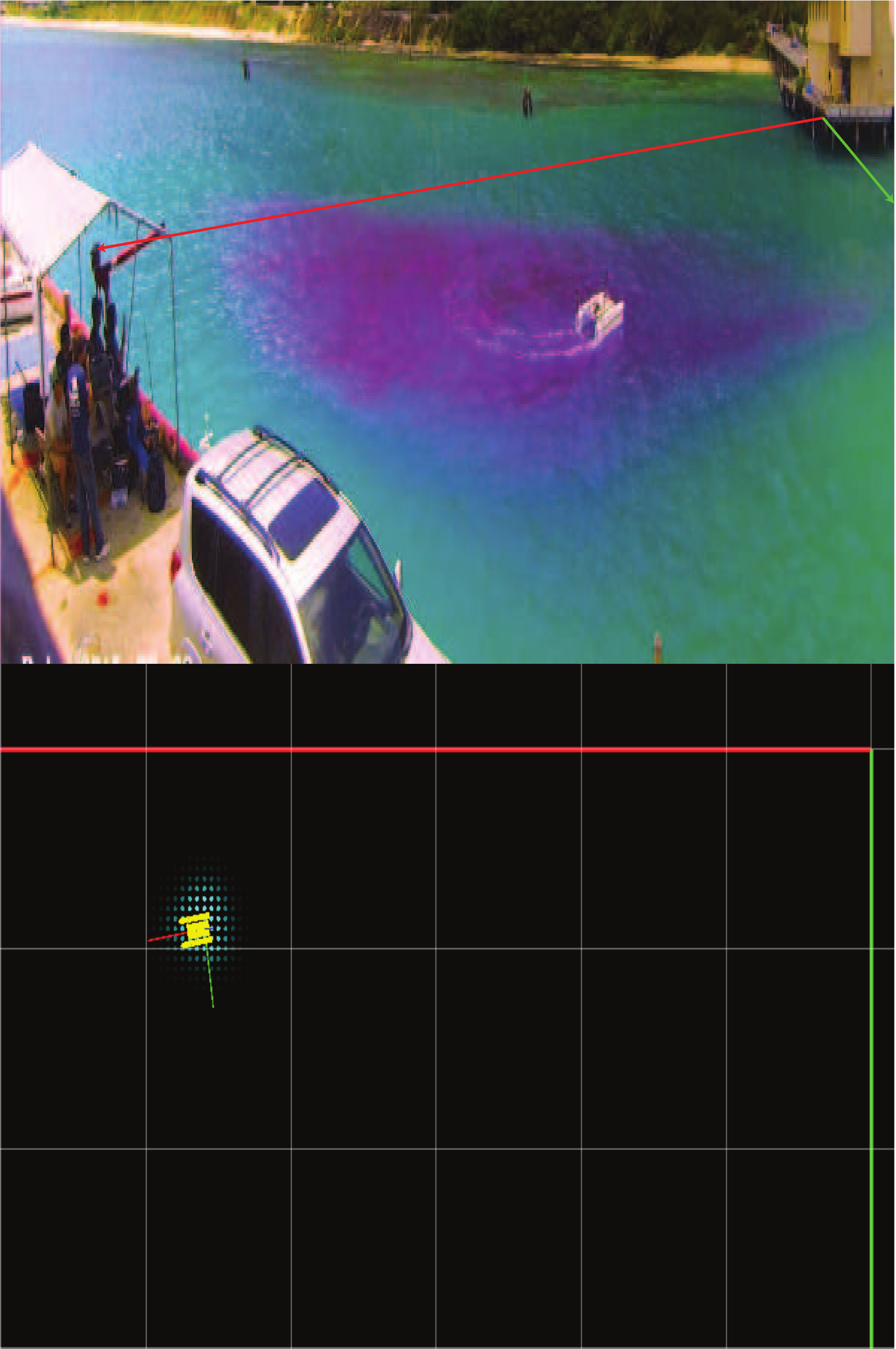}
		\caption{T~=~0~secs}
		\label{fig:50ppb1}
	\end{subfigure}%
	\begin{subfigure}[t]{0.25\textwidth}
		\centering
		\includegraphics[width=0.8\textwidth, height=0.8\textwidth, angle =0]{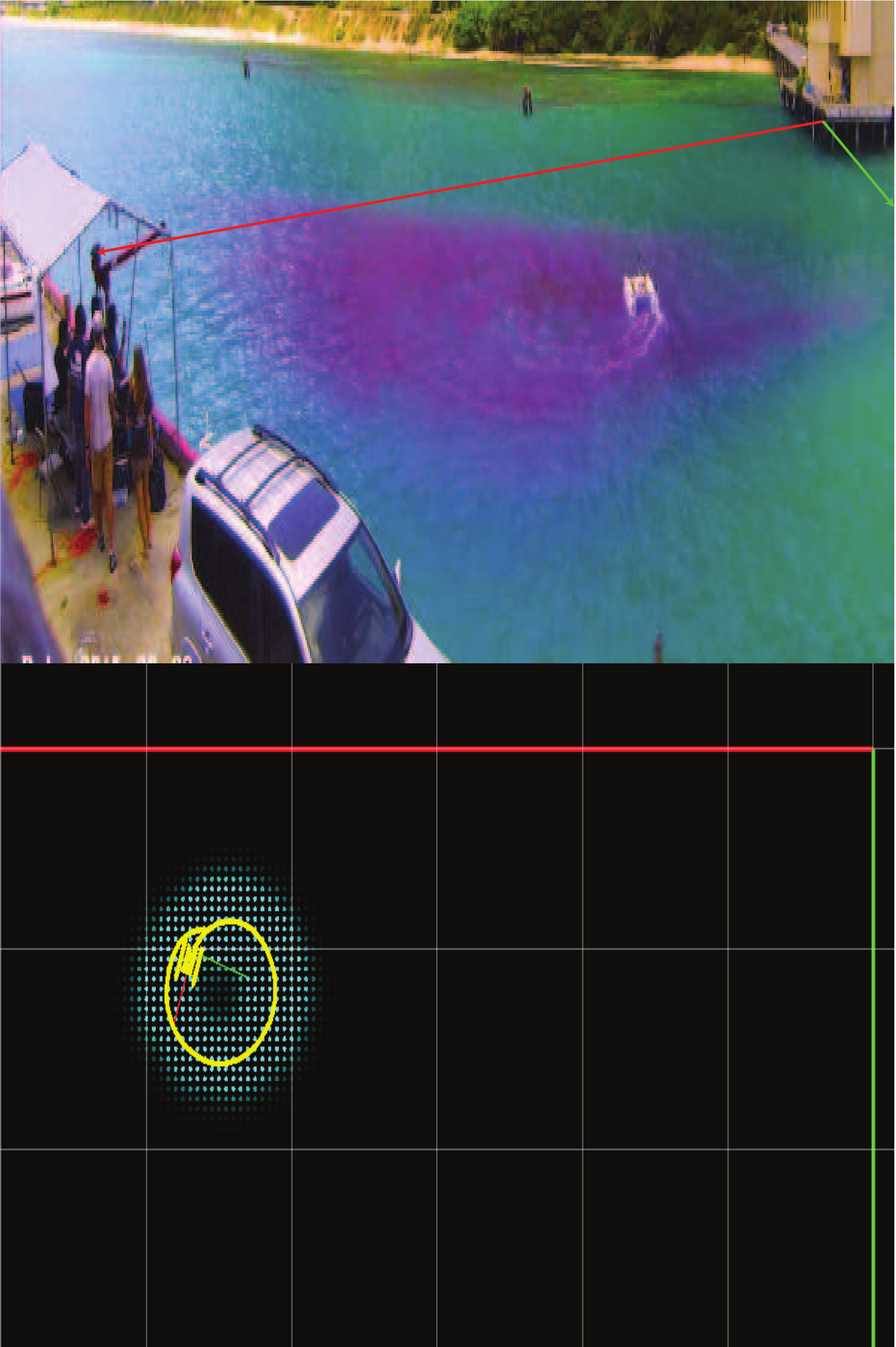}
		\caption{T~=~20~secs}
		\label{fig:50ppb2}
	\end{subfigure}%
	\begin{subfigure}[t]{0.25\textwidth}
		\centering
		\includegraphics[width=0.8\textwidth, height=0.8\textwidth, angle =0]{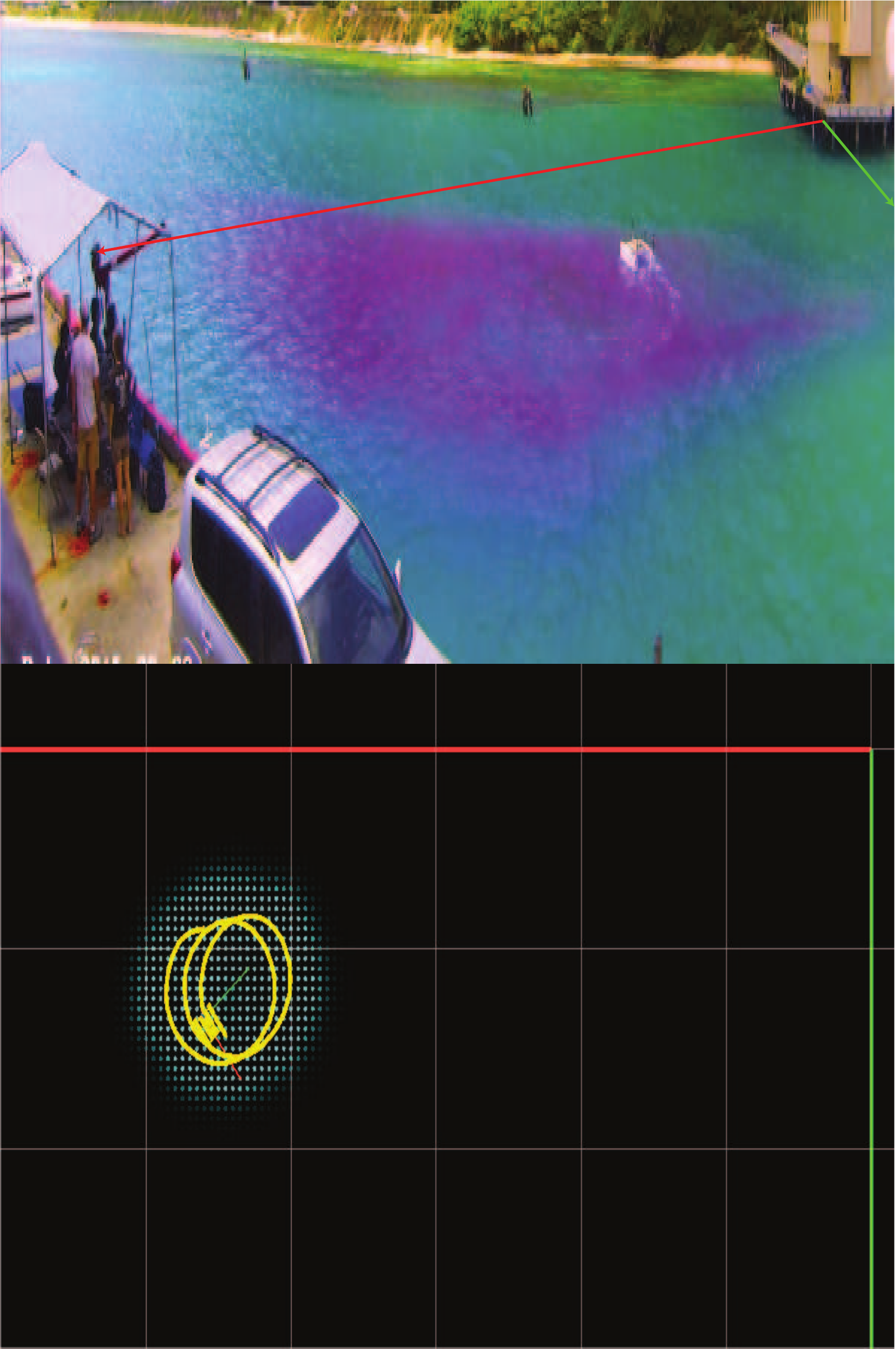}
		\caption{T~=~40~secs}
		\label{fig:50ppb3}
	\end{subfigure}%
	\begin{subfigure}[t]{0.25\textwidth}
		\centering
		\includegraphics[width=0.8\textwidth, height=0.8\textwidth, angle =0]{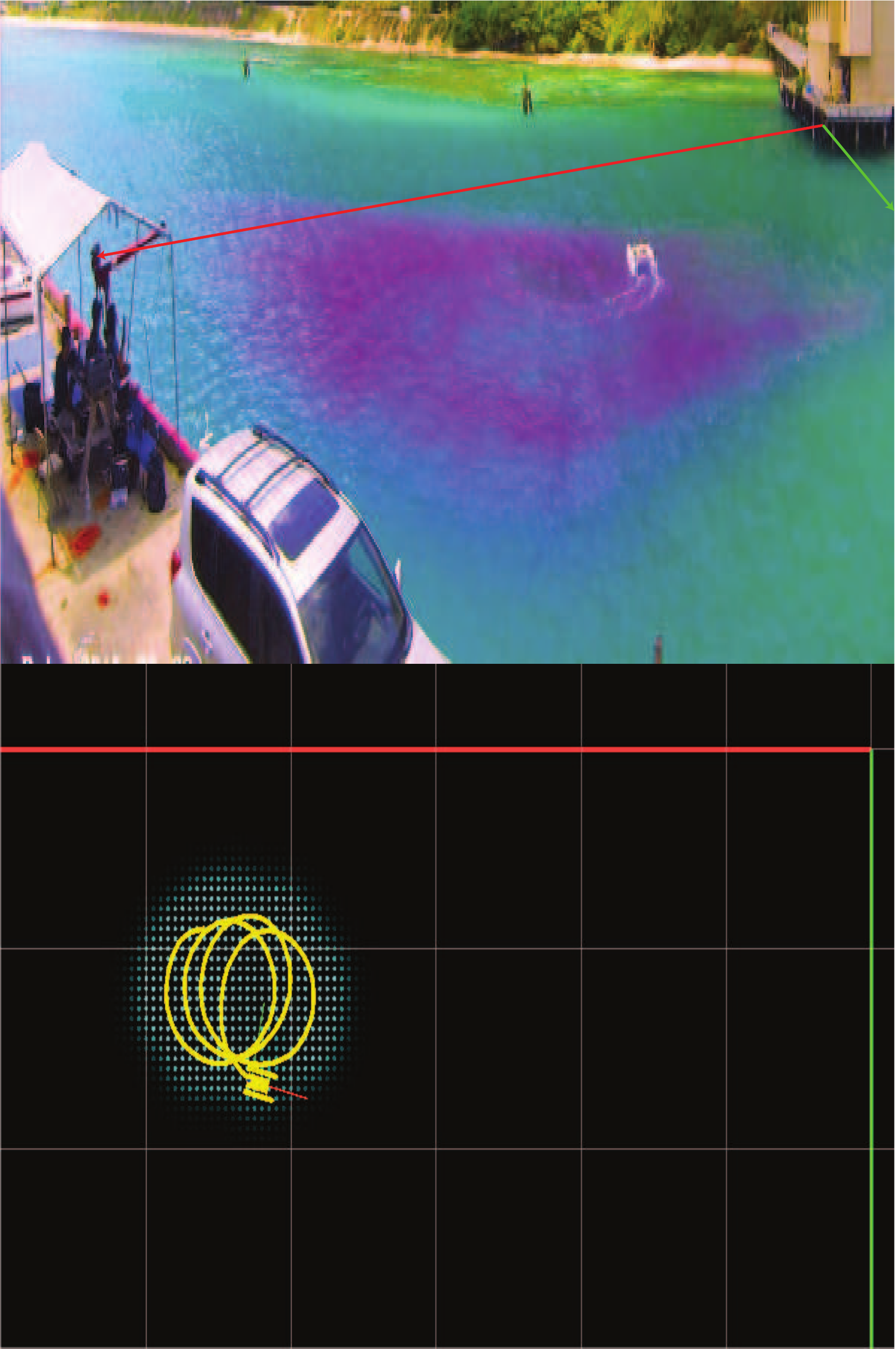}
		\caption{T~=~60~secs}
		\label{fig:50ppb4}
	\end{subfigure}%
	\caption{Trajectory followed by the USV while tracking concentration $c_0$ is set to 50~ppb. The top half of each figure shows the snapshot of the experiments. The bottom half shows the visualization of the experiment in ROS rviz environment.}
	\label{fig:50ppb_tracking_traj}
\end{figure*}

The top pane in Fig.~\ref{fig:50ppb1}-\ref{fig:50ppb4} shows the snap shots of the video recorded during the experiments. It shows the USV, the plume and the approximate location of the x-axis with the red line and y-axes with the blue line. The origin of the local coordinate system is at the edge of the breakwater shown by the intersection point of the two axes. The bottom pane of each figure shows the visualization of the USV, the two axes and plume visualization using the measured concentration from the four sensors in Robot Operating System (ROS) $rviz$ visualization utility. The trajectory followed by the USV is shown by the solid yellow line.
The Fig.~\ref{fig:50ppb1}-\ref{fig:50ppb4} show the USV following a near circular trajectory while tracking the level curve. This circular motion is attributed to the term $\frac{ \mathrm{v}_d  A\nabla c_r }{\|A\nabla c_r\|}$ in (\ref{eqn7}), which makes the USV to patrol the level curve in a counter-clock direction with velocity $\mathrm{v}_d$. The center of the circular trajectory can be seen moving towards the y-axis. This is due to the dynamic nature of the plume, as the environmental conditions cause the plume to advect in that direction. The movement of the USV inside the plume also perturbs the plume due to the strong currents generated by the USV actuators. This can be observed in the sequence of these figures, where the plume is visibly changed within 60 secs, creating a patch of visibly lower concentration towards the center of the plume. 

\begin{figure}[H]
	\centering
	\centering
	\includegraphics[width=0.3\textwidth,height=0.15\textwidth,angle =0]{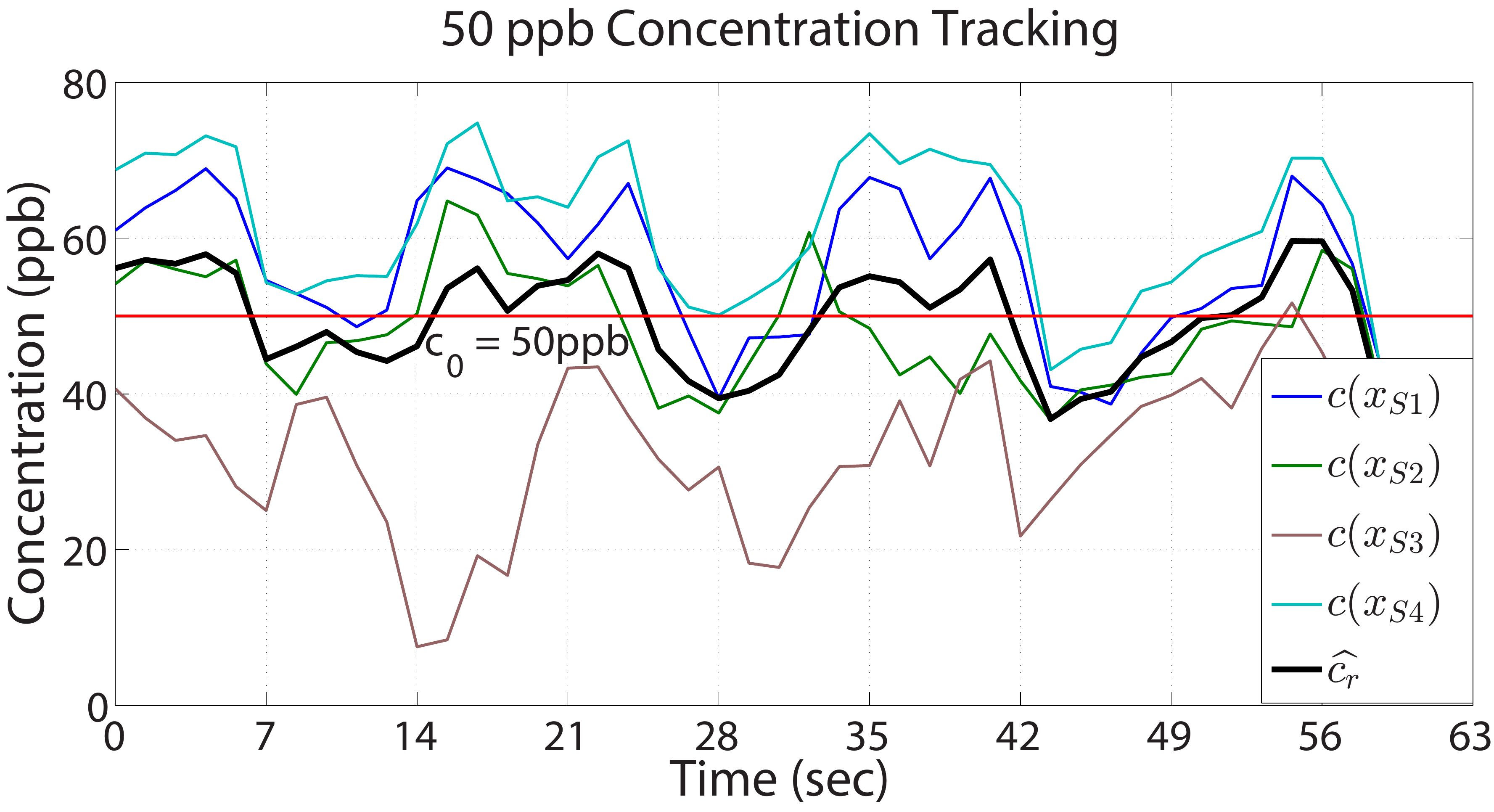}
	\caption{Plot of time series of concentration measurement of four sensors $c(\bm{\mathrm{x}}_{Si})$, $i$=1,2,3,4, the mean of the concentration $\widehat{c_r}$, and the reference line at $c_0$ = 50~ppb.}
	\label{fig:50ppb_tracking}
\end{figure}

The concentration measurements from each sensor on the USV were recorded, in addition to other parameters of the test. These time series of sensor concentration measurements $c(\bm{\mathrm{x}}_{Si})$ for $i$=1,2,3,4, $\widehat{c_r}$ are plotted in Fig.~\ref{fig:50ppb_tracking}. The reference line for tracked concentration $c_0$ at 50~ppb is also shown in Fig.~\ref{fig:50ppb_tracking}. The estimated mean of the concentration $\hat{c_r}$ is shown by the black line shown in the figure. Significant inter-sensor concentration difference can be seen in the Fig.~\ref{fig:50ppb_tracking}, however, the estimated mean concentration value is close to the tracked concentration value $c_0$ set at 50~ppb. 

\subsubsection{Case 2}$\textit{Tracking level concentration curve at 40~ppb}$

The next test was conducted with $c_0$ set to 40~ppb with the same intention as the first test. The values for $k$, $k_1$, $k_2$ and v$_d$ are also the same as those set in Case-1. The top half of Fig.~\ref{fig:40ppb1}-\ref{fig:40ppb4} shows the snapshots of the experiment video. The approximate location of the two axes is also shown. The trajectory followed by the USV is marked by a yellow line and displayed in ROS $rviz$ along with the measured concentration cloud. In this case, the USV again stays within the denser part of the plume towards the center and follows a near circular trajectory as expected. The approximate center of this circular trajectory can be seen advecting towards the x-axis. This direction is different from the last case. We observed during testing that the plume advection direction varies due to changing currents in the testing arena which is the cause of this shift. The perturbation caused by the motion of the USV can also be clearly seen as visibly reduced concentration in the top half of these sequence of figures. This test was also conducted for 60~secs.  

The concentration time series for $c(\bm{\mathrm{x}}_{Si})$ for $i$=1,2,3,4 were recoded and plotted in Fig.~\ref{fig:40ppb_tracking}, which also shows the mean concentration $\widehat{c_r}$ in black color. The reference line for tracked concentration $c_0$ set at 40~ppb is shown in red in this figure. The mean concentration $\widehat{c_r}$ follows the tracked concentration $c_0$, which demonstrates satisfactory performance of the tracking controller. 

\subsection{Discussions}
\begin{figure}[h]
	\centering
	\centering
	\includegraphics[width=0.3\textwidth,height=0.15\textwidth,angle =0]{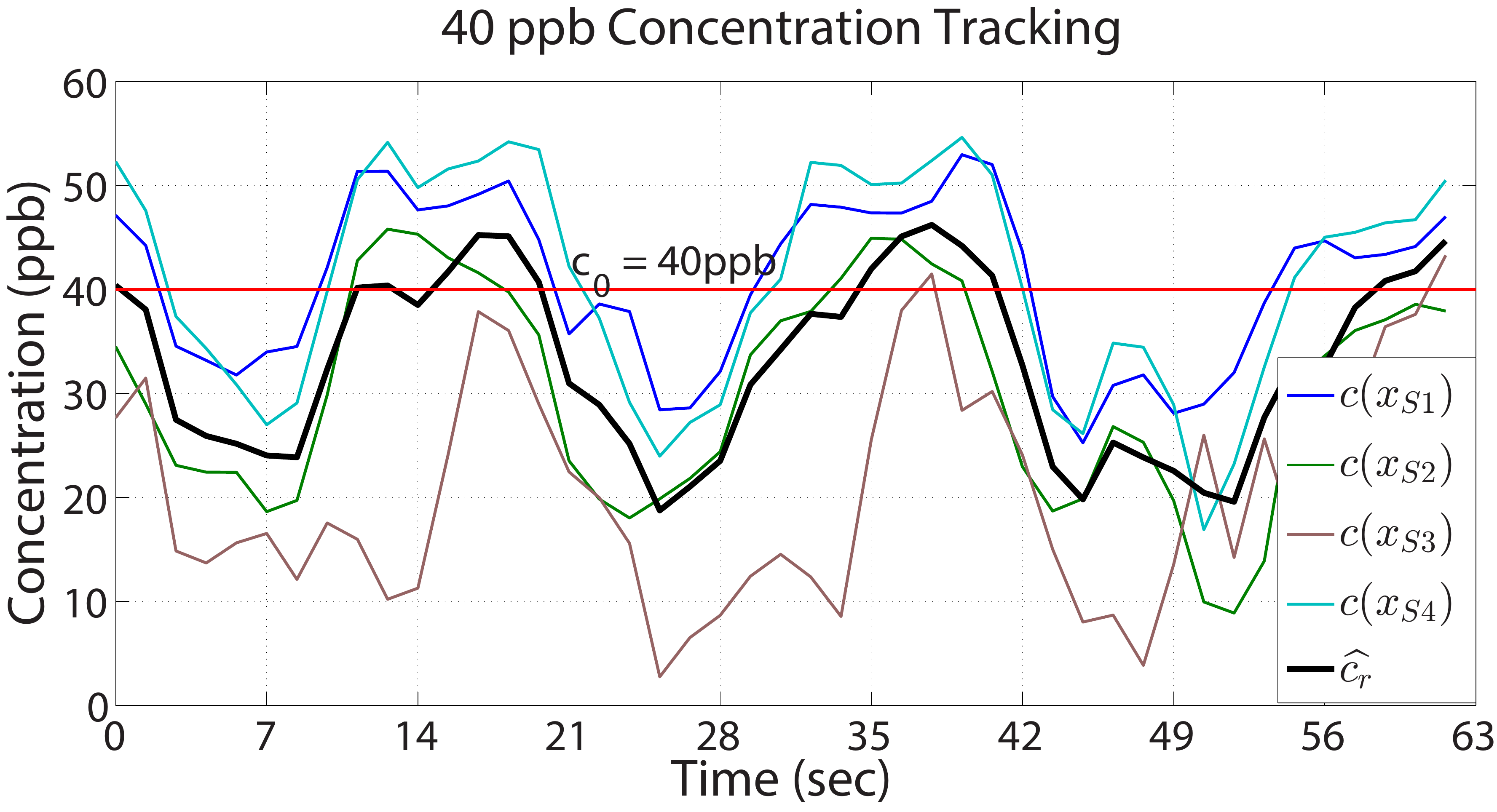}
	\caption{Plot of time series of concentration measurement of four sensors $c(\bm{\mathrm{x}}_{Si})$, $i$=1,2,3,4, the mean of the concentration $\widehat{c_r}$, and the reference line at $c_0$ = 40~ppb.}
	\label{fig:40ppb_tracking}
\end{figure}

Although the field experimental results show that our designed controller works in tracking concentration level curves, there are complexities and limitations involved in the testing of control laws in experimental plumes generated in real marine environments. As we performed the experiments in a near shore environment, the usable size of the testing arena is only of the order 10s of meters. A plume of such a size is relatively small and the concentration characteristics may not represent a larger-scale pollution plume. We think that the small size of the plume results in sharp changes in concentration values and also high gradient values, even over small distance. The concentration time series $c(\bm{\mathrm{x}}_{Si})$ measured by the fluorometer sensors during the plume concentration curve tracking experiments have been plotted in Fig.~\ref{fig:50ppb_tracking} and \ref{fig:40ppb_tracking}. These time series show significant inter-sensor difference in the measured concentration, even though sensor separation is of the order of 2~m or less. This makes plume monitoring more challenging, since even small movements of the USV can cause large changes in the measured concentration, which in turn results in relatively big tracking error.
\begin{figure*}[!]
	\centering
	\captionsetup{justification=centering}
	\begin{subfigure}[t]{0.25\textwidth}
		\centering
		\includegraphics[width=0.8\textwidth, height=0.8\textwidth, angle =0]{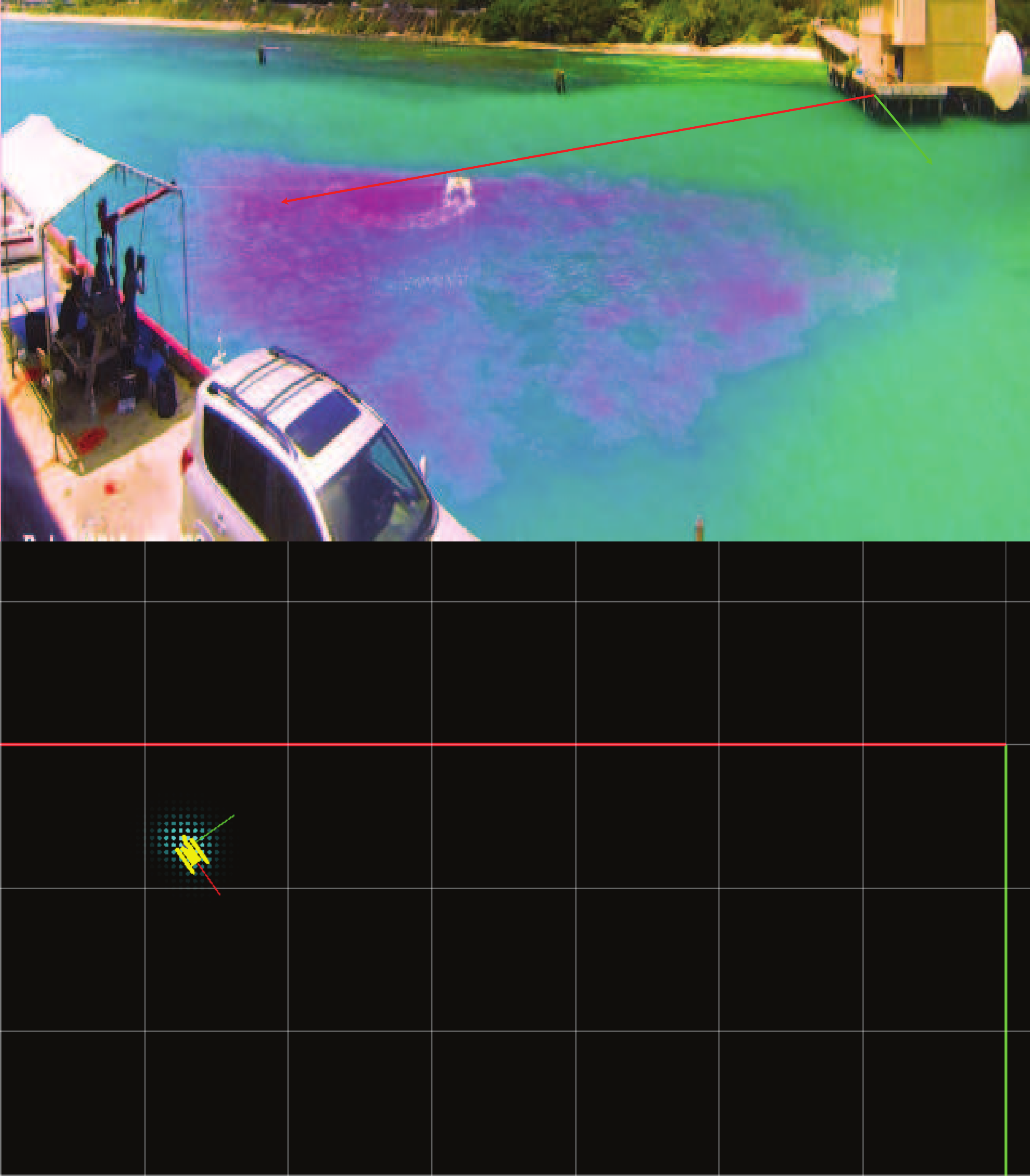}
		\caption{T~=~0~secs}
		\label{fig:40ppb1}
	\end{subfigure}%
	\begin{subfigure}[t]{0.25\textwidth}
		\centering
		\includegraphics[width=0.8\textwidth, height=0.8\textwidth, angle =0]{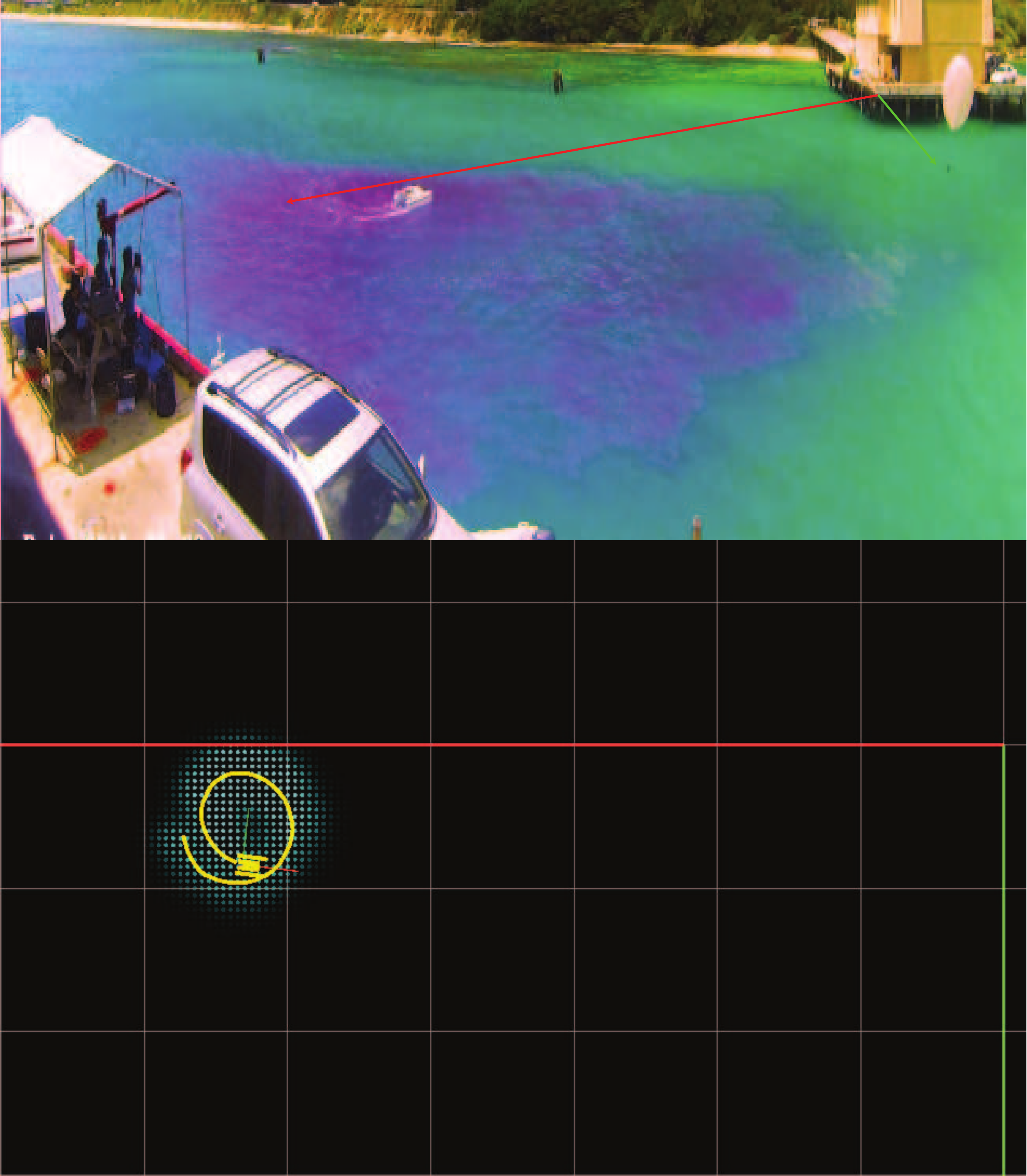}
		\caption{T~=~20~secs}
		\label{fig:40ppb2}
	\end{subfigure}%
	\begin{subfigure}[t]{0.25\textwidth}
		\centering
		\includegraphics[width=0.8\textwidth, height=0.8\textwidth, angle =0]{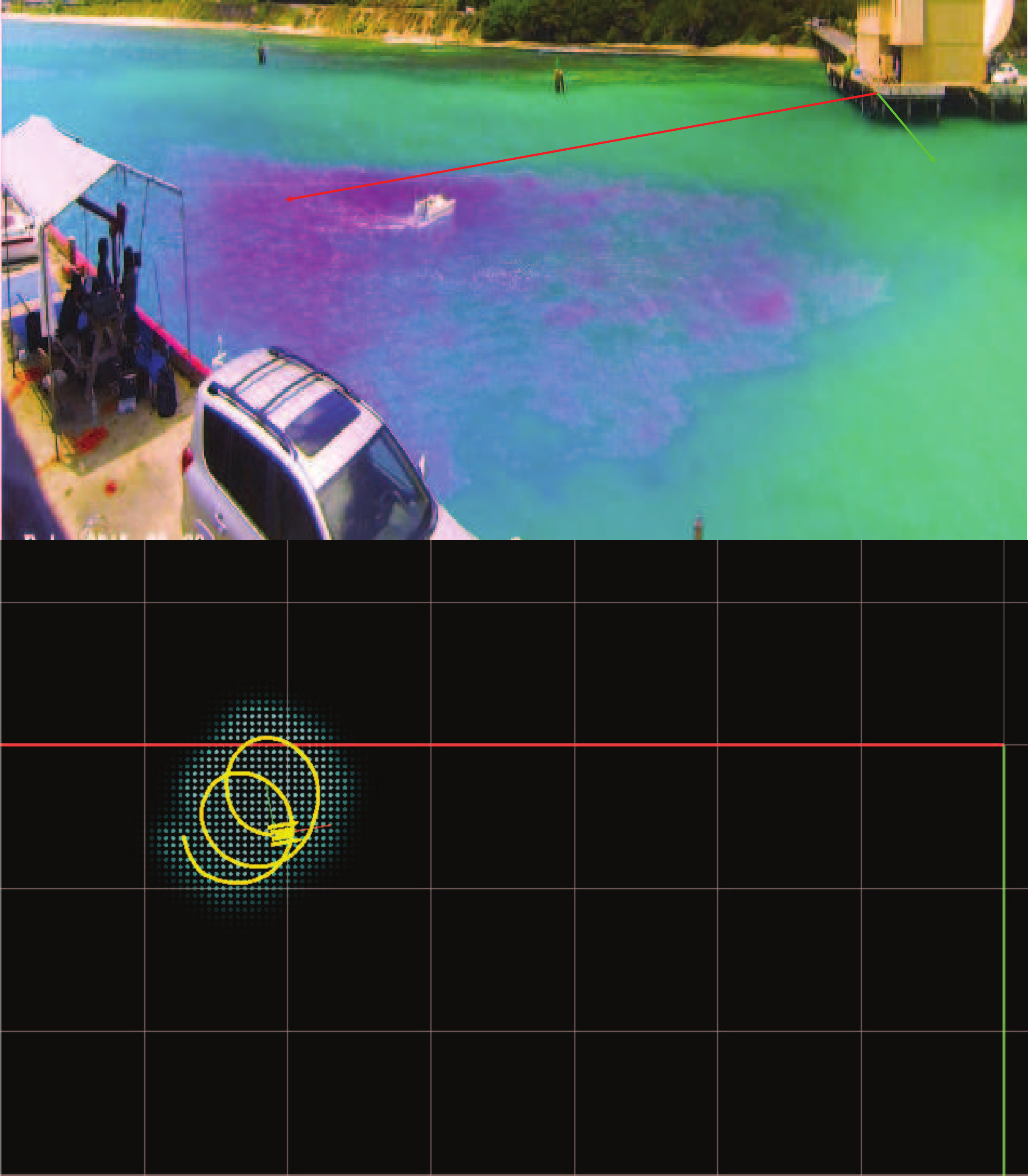}
		\caption{T~=~40~secs}
		\label{fig:40ppb3}
	\end{subfigure}%
	\begin{subfigure}[t]{0.25\textwidth}
		\centering
		\includegraphics[width=0.8\textwidth, height=0.8\textwidth, angle =0]{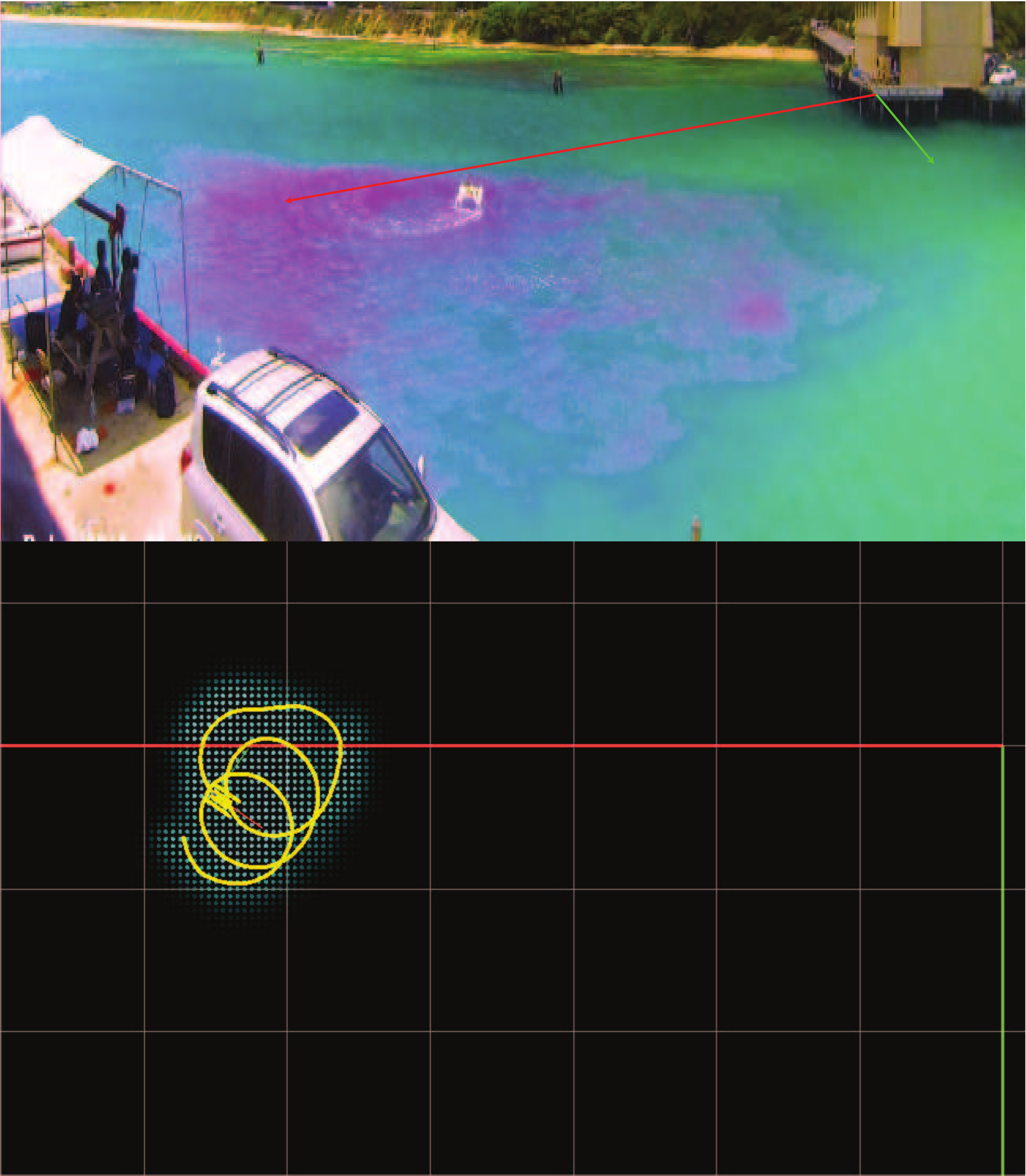}
		\caption{T~=~60~secs}
		\label{fig:40ppb4}
	\end{subfigure}%
	\caption{Trajectory followed by the USV while tracking concentration $c_0$ is set to 40~ppb. The top half of each figure shows the snapshot of the experiments. The bottom half shows the visualization of the experiment in ROS rviz environment.}
	\label{fig:40ppb_tracking_traj}
\end{figure*}
Also, the motion of the USV perturbs the plume. This perturbation caused the plume to develop differently than the ideal model presented in (\ref{eqn1}). Since the scale of the plume is relatively small, the effect of this perturbation significantly alters the plume structure. Moreover, the longer the USV moves inside the plume, the more significant the effect of this perturbation. This can be seen in Fig.~\ref{fig:50ppb_tracking_traj} and \ref{fig:40ppb_tracking_traj}, where the shape of the plume is visibly changed within 60~secs of the experiment. The circular motion of the USV generates a region of low concentration at the center of the USV trajectory. This further complicates the plume monitoring experiments.

\section{Conclusion and Future Work}\label{sec:conclusion}

In this paper, we presented a  a control design to track concentration level curves of a pollution plume dispersion using a USV. The presented algorithms were tested in field experiments using a USV monitoring Rhodamine dye plumes. Two test cases are presented to demonstrate the performance of the proposed control law. The controller performed reasonably well despite the challenges presented in real field experiments such as high concentration gradients and the perturbations caused by the motion of the USV in the plume. Videos of our experimental work can be found at our Vimeo site\footnote{Plume Tracking Field Experiment Results at \url{https://vimeo.com/user45850047}}. Using the results gathered during these experiments, we plan to develop a more robust, behavioral/gradient-based hybrid controller for plume tracking, in oder to overcome some of the shortcomings of the current controller design and to improve the tracking performance.

\bibliographystyle{ieeetr}
\bibliography{journal_2016}
\end{document}